\documentclass[aps,prd,twocolumn,showpacs,nofootinbib,eqsecnum,superscriptaddress]{revtex4-1}

\usepackage[centertags]{amsmath}
\usepackage{amssymb}
\usepackage{latexsym}
\usepackage{enumerate}
\usepackage{graphicx}
\usepackage{mathrsfs}
\usepackage[hidelinks]{hyperref}
\usepackage{stmaryrd}
\usepackage{import}
\usepackage{tensor}
\usepackage{xcolor}
\usepackage{bm}
\usepackage{multirow}
\usepackage{breakurl}
\usepackage{float}
\usepackage{lipsum}
\usepackage{siunitx}
\usepackage{comment}
\usepackage{mathtools}
\usepackage{autobreak}
\usepackage[caption=false]{subfig}
\usepackage{orcidlink}

\mathtoolsset{centercolon}

\usepackage{stmaryrd}
\usepackage{color}
\usepackage{amsmath}
\usepackage{amssymb}
\usepackage{latexsym}
\usepackage{enumerate}
\usepackage{graphicx}
\usepackage{mathrsfs}
\usepackage{physics}
\usepackage{tabularx}
\usepackage{mathtools}
\makeindex

\newcommand{\be}{\begin{equation}}
\newcommand{\ee}{\end{equation}}
\newcommand{\bea}{\begin{eqnarray}}
\newcommand{\eea}{\end{eqnarray}}
\newcommand{\ba}{\begin{array}}
\newcommand{\ea}{\end{array}}
\newcommand{\bi}{\begin{itemize}}
\newcommand{\ei}{\end{itemize}}

\newcommand{\D}{\Delta}

\newcommand{\la}{\lambda}
\newcommand{\vp}{\varphi}
\renewcommand{\O}{\Omega}

\renewcommand{\th}{\theta}

\newcommand{\Up}{\Upsilon}

\newcommand{\atil}{\tilde{a}}

\newcommand{\coeF}[2]{\mathcal{#1}_{#2}}
\newcommand{\Gam}{\Gamma}

\allowdisplaybreaks

\begin{document}

\title{Post-Newtonian Expansion of Fluxes from a Scalar Charge on an Inclined-Spherical Orbit about a Kerr Black Hole}

\begin{abstract}
    Efforts are underway to accurately model extreme-mass-ratio inspirals for binaries with a spinning (Kerr) primary. At lowest order the adiabatic evolution depends on the radiation fluxes.  Fluxes and other self-force quantities can be expanded analytically in post-Newtonian (PN) series allowing the early evolutionary phase to be understood.  When it comes to more complicated background geodesic orbits, it proves useful to use the scalar field model problem to guide development and testing of techniques.  In this paper, we present analytical expressions for the scalar fluxes from a scalar point-charge in inclined-spherical orbit about a Kerr black hole up to 12PN relative order, with expressions that are exact in terms of the inclination parameter $x$ and black hole spin $a$. The expressions are constructed using the Mano, Suzuki, and Takasugi method of solving the scalar wave equation in a Kerr background. We compare the numerical evaluation of these flux expressions to full numerical ($s=0$) Teukolsky code results, examining their degree of utility as the strong-field region is approached.
\end{abstract}

\author{Jezreel C. Castillo\,\orcidlink{0000-0001-6617-0326}}
\affiliation{Department of Physics and Astronomy, University of North Carolina at Chapel Hill, Chapel Hill, North Carolina 27599}
\author{Charles R. Evans\,\orcidlink{0000-0001-5578-1033}}
\affiliation{Department of Physics and Astronomy, University of North Carolina at Chapel Hill, Chapel Hill, North Carolina 27599}
\affiliation{School of Mathematics and Statistics, University College Dublin, Belfield D04 N2E5, Dublin 4, Ireland}
\author{Chris Kavanagh\,\orcidlink{0000-0002-2874-9780}}
\affiliation{School of Mathematics and Statistics, University College Dublin, Belfield D04 N2E5, Dublin 4, Ireland}
\author{Jakob Neef\,\orcidlink{0000-0002-3215-5694}}
\affiliation{School of Mathematics and Statistics, University College Dublin, Belfield D04 N2E5, Dublin 4, Ireland}
\author{Barry Wardell\,\orcidlink{0000-0001-6176-9006}}
\affiliation{School of Mathematics and Statistics, University College Dublin, Belfield D04 N2E5, Dublin 4, Ireland}
\author{Adrian Ottewill\,\orcidlink{0000-0003-3293-8450}}
\affiliation{School of Mathematics and Statistics, University College Dublin, Belfield D04 N2E5, Dublin 4, Ireland}

\maketitle

\section{Introduction}

With the Laser Interferometer Space Antenna (LISA) mission having secured funding from the European Space Agency, theoretical work continues on modeling likely LISA gravitational-wave sources \cite{AmarETC07}.  Extreme-mass-ratio inspirals (EMRIs) are of particular interest: binary systems involving a compact object of mass $\mu$ orbiting a massive central black hole of mass $M$, with mass ratios of order $\varepsilon = \mu/M \sim10^{-6}-10^{-4}$.  EMRI orbits are expected to be inclined and eccentric, and to have a gradual adiabatic evolution through the LISA passband that allows high-precision parameter estimation \cite{GairETC17,ChapETC25}.

Due to their small mass ratios, $\varepsilon \ll 1$, EMRIs are best modeled using black hole perturbation theory. The overall approach is called self-force theory \cite{Bara09,PoisPounVega11}, where the EMRI is modeled as a compact object perturbing the background spacetime of a rotating (Kerr) black hole and generating a self-force (radiation reaction or tail field) that affects its own motion.  The goal is to accurately calculate both the adiabatic influence and the first post-adiabatic correction \cite{WardETC23}, which is necessary for comparison to LISA data.   

Full EMRI models will involve eccentric and inclined orbits in the self-force calculation. While LIGO-Virgo-KAGRA binaries have little-to-no eccentricity due to circularization \cite{Pete64}, EMRIs are expected to arrive in LISA's passband with moderate eccentricities \cite{GlamKenn02,HopmAlex05}. Inclination will generally be important too, as it exhibits, if anything, mild increases as a binary spirals inward \cite{Hugh00b,DellETC24}.
    
Another avenue for modeling EMRIs involves a synthesis between post-Newtonian (PN) theory and self-force theory (SF-PN), made possible by analytically PN-expanding solutions to the Teukolsky equation.  For low-multipoles, the approach uses the method of Mano, Suzuki and Takasugi (MST) \cite{ManoSuzuTaka96a,ManoSuzuTaka96b}.  This hybrid method has allowed high-order PN expansions of fluxes \cite{ForsEvanHopp16,MunnETC20,MunnEvanFors23, SagoFujiNaka25,CastETC25a} and conservative self-force quantities \cite{KavaOtteWard15,KavaOtteWard16,MunnEvan22a,MunnEvan22b}.

This paper is part of a series on calculating SF-PN expansions for the interesting special case of Kerr-EMRIs with inclined spherical orbits.  Recently, we presented analytic expressions of gravitational fluxes from inclined spherical orbits \cite{CastETC25a} (simultaneously with the appearance of an independent and similar computation \cite{SagoFujiNaka25}). There we computed the fluxes to infinity and the horizon up to 12PN relative order, with expansions that are exact functions of spin $a$ and the inclination parameter $x$ at each PN order. The PN expanded fluxes were validated at large orbital separations by numerical results from a Teukolsky code, with the comparison further used to examine how useful the PN expansions might be for tight orbits.    

Several next steps seem obvious.  One task would be to extend the analytical calculation to the {\it conservative} portion of the gravitational self-force for inclined orbits, requiring an interesting exploration of extending the mode-sum regularization procedure.  We hope to return to that issue in a future paper.  A second task, which is the subject of this paper, is to apply the method of PN expanding fluxes to a simpler model--that of a scalar charged particle in an inclined-spherical orbit, radiating scalar waves.  The scalar model problem has often served as a simple way to develop new techniques prior to extension to the gravitational case, such as the original understanding of the self-force on point-particles \cite{Quin00}; radiation reaction from Kerr circular orbits \cite{GralFrieWise05}; numerical self-force in Kerr circular orbits \cite{WarbBara10}, Kerr eccentric equatorial orbits \cite{WarbBara11}, Kerr inclined spherical orbits \cite{Warb15}, and Kerr generic orbits \cite{NasiOsbuEvan19}; and self force on particles orbiting Reissner-Nordstrom black holes \cite{BiniCarvGera16}.  It might provide a way of exploring some of the issues associated with PN expanding the conservative self-force in inclined Kerr-EMRIs, specifically how to handle analytically the sums over $m$ modes.  In addition, an analytical expression for the scalar field may be of interest for ongoing research on EMRIs as probes of scalar fields \cite{MaseETC21,BarsETC22a,BarsETC22b,DellETC24,SperETC24}.  
    
For these reasons at least, it seems worthwhile to consider the scalar field problem and to first obtain the expansions of the dissipative part of the scalar self-force.  Hence, similar to our previous work \cite{CastETC25a}, in this paper we compute the energy and angular momentum fluxes, both to infinity and down the horizon, from a scalar-charged particle on an inclined-spherical orbit about a Kerr black hole.  We again give explicit results up to 12PN relative order and are able to find SF-PN expressions that exhibit closed-form analytic dependence in $a$ and $x$ at each PN order.         
    
This paper is organized as follows.  We first give a brief discussion in Sec.~II of the properties of inclined spherical orbits about a Kerr black hole.  Sec. III outlines the formalism for treating the scalar field generated by a point-mass carrying a scalar charge.  There we discuss the techniques used to analytically solve for modes of the scalar field (i.e., the MST method) and obtain their subsequent PN expansions.  The discussion includes both how to handle the radial functions and the re-expansion of (spin-0) spheroidal harmonics in terms of spherical harmonics, and the PN truncation of that resulting series.  In Sec. IV we provide our analytic results for the PN-expanded scalar fluxes at infinity, complete through 5PN order and highlight a few special higher-order terms.  The remaining terms can be found in the complete 12PN expansion provided in the \texttt{PostNewtonianSelfForce} package of the Black Hole Perturbation Toolkit \cite{BHPTK18,PostNewtonianSelfForce} and in a separate repository \cite{UNCGrav22}.  Sec.~IV concludes with a numerical comparison between our evaluated expressions for specific orbits and corresponding data from a full numerical scalar flux (Teukolsky $s=0$) code.  Since the expressions for the fluxes on the horizon are similar, albeit even more complicated, we relegate the discussion of those terms to an Appendix.  As with the infinity-side fluxes, the complete 12PN (relative) horizon flux expansions are provided in the online repositories.  Throughout this paper we use the $(-,+,+,+)$ metric signature and geometrized units, so that $G = 1$ and $c = 1$.

\section{Spherical Orbits in Kerr Spacetime} \label{sec:geodesics}

Since the geodesic motion we consider here, inclined spherical orbits, is the same as that in our previous paper \cite{CastETC25a}, we provide only a brief summary.  The reader interested in further 
details is referred to that prior paper and references therein.  In Kerr spacetime, in Boyer-Lindquist coordinates, timelike geodesic motion about a black hole of mass $M$ and spin $a$ is governed by the following general set of equations \cite{Cart68,Schm02,DrasHugh06}
    \begin{align}
    \left(\frac{dr}{d\lambda} \right)^2
    &= \left[\mathcal{E} \varpi^2 - a\mathcal{L}_z \right]^2
    -\D \left[r^2+(\mathcal{L}_z-a \mathcal{E} )^2+Q\right] \notag \\
    &\equiv V_r(r) , \label{eq:R} \\
    \left(\frac{d\th}{d\lambda}\right)^2
    &= Q-\mathcal{L}_z^2 \cot^2 \th -a^2(1-\mathcal{E} ^2) \cos^2\th \notag \\
    &\equiv V_\theta(\th) ,  \label{eq:Th}\\
    \frac{d\vp}{d\lambda} &= \Psi^{(r)}(r)+\Psi^{(\th)}(\th) - a\mathcal{E}, \label{eq:Ph}\\
    \frac{dt}{d\lambda} &= T^{(r)}(r)+T^{(\th)}(\th) + a\mathcal{L}_z, \label{eq:T}
    \end{align}
where $\varpi^2 = r^2+a^2$ and $\D = r^2 - 2 M r + a^2$.  From the symmetry of Kerr spacetime there are conserved quantities; $\mathcal{E}$ is the specific energy, $\mathcal{L}$ is the specific angular momentum, $Q$ is the Carter constant.  Additionally, $\lambda$ is Mino time \cite{Mino03,FujiHiki09}, which can be related to proper time $\tau$ by integrating $d\tau/d\lambda = \Sigma = r^2 +a^2 \cos^2\theta$.  The four functions found in the evolution of $t$ and $\vp$ are
\begin{align}
    \label{eqn:PhiR}
    &\Psi^{(r)}(r) = a \mathcal{E} \frac{\varpi^2}{\D}
    -\frac{a^2\mathcal{L}_z}{\D} , \\
    \label{eqn:PhiTh}
    &\Psi^{(\th)}(\th) = \mathcal{L}_z \csc^2\th , \\
    \label{eqn:TR}
    &T^{(r)}(r) = \mathcal{E} \frac{\varpi^4}{\D}
    - a \mathcal{L}_z\frac{\varpi^2}{\D} , \\
    \label{eqn:TTh}
    &T^{(\th)}(\th) = -a^2 \mathcal{E}\sin^2\th .
\end{align}

While any timelike geodesic is fully characterized by the conserved quantities $\{\mathcal{E},\mathcal{L}_z,\mathcal{Q}\}$, it is conventional to make a one-to-one correspondence with an alternative parameter set $\{p,e,x\}$, representing the semi-latus rectum, eccentricity, and inclination parameter, respectively.  This latter set makes the restriction to inclined spherical ($e=0$) orbits more immediate.  Spherical orbits are thus described by just the two parameters, $p$ and $x$, and simultaneously require $V_{r}(p) = 0$ and $V_{r}'(p) =0$, making the radial behavior trivial ($r = r_0 = M p$).  These lead to two conditions on $\mathcal{E}$, $\mathcal{L}_z$, and $\mathcal{Q}$.

The inclination parameter is defined as $x = \sin\th_\mathrm{min}$, where $\th_\mathrm{min}$ is a turning 
point of polar motion found by setting $V_{\th}(\th_{\rm min}) = 0$.  Use of $x$ differs from an alternative inclination angle, $\iota$, that is also frequently used \cite{Hugh01,GanzETC07,Warb15}.  This last condition determines a third relationship between $\mathcal{E}$, $\mathcal{L}_z$, and $\mathcal{Q}$, allowing each to be found in principle as a function of $x$ and $p$.  For our purposes, $1/p$ serves as a natural (gauge specific) post-Newtonian expansion parameter.  It is then easy to separate and write out PN expansions in $1/p$ (or more correctly $1/p^{1/2}$) for $\mathcal{E}$, $\mathcal{L}_z$, and $\mathcal{Q}$, which are also (exact) functions of $a$ and $x$ at each PN order.  The leading part of these expansions can be found in Eqs.~(2.12), (2.13), and (2.14) in \cite{CastETC25a}.

The polar motion in $\theta$ is typically reparameterized with the Darwin angle $\chi$ 
\cite{Darw59,Darw61,DrasHugh04,Warb15}
\begin{align}
    \cos\th(\chi) = \sqrt{1-x^2}\cos\chi , 
    \label{eq:thfunc}
\end{align}
making obvious the turning points in $\th$.  Use of $\chi$ for the polar motion then requires us to find Mino time $\lambda$ as a function of $\chi$.  This relationship follows from integrating \cite{FujiHiki09} the differential equation
\begin{align}
\label{eqn:dchidla}
    \frac{d\chi}{d\lambda} = \sqrt{a^2(1-\mathcal{E}^2)(z_+-z_-\cos^2\chi)},
\end{align}
where $z_- = (1-x^2)$ and $z_+ = \mathcal{Q}/(a^2(1-\mathcal{E}^2)(1-x^2))$, which turn out to be the roots of $V_{\theta}(\th)$ with $z=\cos^2\th$.  

With the motion in $\th$ (or $\chi$) determined as a function of $\lambda$, the evolution of coordinate time $t$ and azimuth angle $\vp$ can then likewise be obtained from Eqs.~\eqref{eq:Ph} and \eqref{eq:T} as quadratures over $\lambda$.  Each has both a linear (cumulative) and a periodic dependence
\begin{align}
    \label{eqn:tpOfLa}
    t_p(\la) &= \Upsilon_t  \la + \D t^{(\th)}(\la) ,
    \\
    \label{eqn:phipOfLa}
    \vp_p(\la) &= \Upsilon_{\vp} \la + \D \vp^{(\th)}(\la) ,
\end{align}
where $\Up_t$ and $\Up_{\vp}$ are the mean rates of advance of $t$ and $\vp$, respectively.  The behavior of 
Eqs.~\eqref{eqn:tpOfLa} and \eqref{eqn:phipOfLa}, as well as the oscillation from \eqref{eqn:dchidla}, lead to three Mino-time frequencies (or rates), $\Up_\th,\Up_\vp,\Up_t$ \cite{FujiHiki09}.  These in turn determine two fundamental frequencies with respect to Boyer-Lindquist time
\begin{align}
    \O_\th = \frac{d\th}{dt} = \frac{\Up_\th}{\Up_t}, \qquad \O_\vp = \frac{d\vp}{dt} = \frac{\Up_\vp}{\Up_t}.
\end{align}
See \cite{CastETC25a} for more details.

Inclined spherical orbits are therefore biperiodic.  Connecting the scalar field to the source motion requires knowing the functional dependencies of $t(\lambda)$ and $\vp(\lambda)$ (or the equivalent functions of $\chi$), which are expressed as double Fourier series.  In this way, the scalar field becomes decomposed into a doubly infinite set of modes over harmonics of the two fundamental frequencies.  These frequencies, and the Fourier series for the motion, are obtained as lengthy analytic expressions once we make a PN expansion.  Leading terms in the PN expansion of the frequencies can be found in Sec.~II of \cite{CastETC25a}.

\section{The Scalar Field} \label{sec:scalars}

\subsection{Scalar Wave Equation in Kerr Spacetime}

The scalar field is governed by the massless, minimally coupled Klein-Gordon equation 
\begin{align}
    \square\Phi = -4\pi T ,
\end{align}
where $T$ in our application is a scalar point-charge density, defined in what follows.  In Kerr geometry and Boyer-Lindquist coordinates, the wave operator can be expanded \cite{BrilETC72,Teuk73} out, leading to the field equation
\begin{widetext}
\begin{align}
    \frac{\partial}{\partial r}\left(\Delta\frac{\partial\Phi}{\partial r}\right) +\left(\frac{1}{\sin^2\theta}- \frac{a^2}{\Delta}\right)\frac{\partial^2\Phi}{\partial\vp^2} - \frac{4 M a r}{\Delta}\frac{\partial^2\Phi}{\partial\vp\partial t}-\left(\frac{\varpi^4}{\Delta}-a^2\sin^2\theta\right)\frac{\partial^2\Phi}{\partial t^2} + \frac{1}{\sin\theta}\frac{\partial}{\partial\th}\left(\sin\theta\frac{\partial\Phi}{\partial\theta}\right) = -4 \pi \Sigma \, T .
\end{align}
As is long known \cite{Cart68,BrilETC72}, this equation is separable in the frequency domain in terms of spheroidal harmonics
\begin{align}
\label{eqn:Decomp} 
\Phi = \frac{1}{2 \pi} \int \sum_{\ell m} R_{\ell m\omega}(r)\, S_{\ell m}(\theta ;\sigma^2)\, e^{-i\omega t+im\vp}\, d\omega , 
\qquad \qquad
\Sigma \, T = \frac{1}{2 \pi} \int\sum_{\ell m} T_{\ell m\omega}(r)\, S_{\ell m}(\theta ;\sigma^2)\, e^{-i\omega t+im\vp}\, d\omega . 
\end{align}
Here, $S_{\ell m}(\theta ;\sigma^2)$ are the spheroidal Legendre functions with spheroidicity $\sigma^2 = -a^2 \omega^2$.  Substitution of \eqref{eqn:Decomp} into the field equation then leads to the separate angular and radial equations
\begin{align}
    &\frac{1}{\sin\th}\frac{d}{d\th}\left(\sin\th \frac{d S_{\ell m \omega}}{d\th}\right) 
   + \left(\lambda_{\ell m \omega} +a^2\omega^2 \cos^2\th -\frac{m^2}{\sin^2\th}\right) S_{\ell m \omega} = 0 ,
    \label{eq:spheroidH} 
    \\
    &\Delta \frac{d}{d r}\left(\Delta\frac{d R_{\ell m \omega}}{d r}\right) 
    + \left(m^2 a^2 - 4 M a m \omega r + (r^2 + a^2)^2 \omega ^2 - \omega^2 a^2 \Delta 
    - \lambda_{\ell m \omega} \Delta \right) R_{\ell m \omega} = -4 \pi \Delta \, T_{\ell m \omega}. 
    \label{eq:radTeukEq}
\end{align}
where $S_{\ell m \omega} = S_{\ell m}(\th ; -a^2 \omega^2)$ and $\lambda_{\ell m \omega}$ are the eigenvalues that insure regularity of solutions of \eqref{eq:spheroidH} at $\th = 0,\pi$ \cite{CasaOtte05}.  The spheroidal harmonics are then the combinations, $S_{\ell m}(\th ;\sigma^2) e^{i m \vp}$, which are orthogonal and normalized such that
\begin{equation}
    \int S_{\ell m}(\th ;\sigma^2) e^{i m \vp} S_{\ell' m'}(\th ;\sigma^2) e^{-i m' \vp} \, d\Omega 
    = \delta_{\ell \ell'} \, \delta_{m m'} .
\end{equation}
The radial equation is identical to the spin-0 Teukolsky equation \cite{Teuk73}, allowing us to apply the MST formalism. 

\subsection{Solutions to the scalar wave equation with a point-charge source}

For a scalar point-charge on an inclined spherical orbit, the source function can be written as 
\begin{align}
    T = q \int \delta^{(4)}(x^\mu - x^\mu_p (\tau)) (-g)^{-1/2} \, d\tau 
      = \frac{q}{\Sigma \sin\th \, u^t(\th_p (t))} \delta(r - r_0)\, \delta(\th - \th_p (t)) \, \delta(\vp - \vp_p (t)) ,
\end{align}
\end{widetext}
which can be equated to the frequency-domain mode decomposition of $T$ found in \eqref{eqn:Decomp}.  Then, the biperiodicity of the angular motion, captured by the delta functions, gives rise to a discrete frequency spectrum consisting of harmonics of the fundamental frequencies
\begin{align}
    \omega = \omega_{mk} = m\Omega_\vp+k\Omega_\theta .
\end{align}
The Fourier transform integral in \eqref{eqn:Decomp} for $T$ is replaced by a third sum,
\begin{align}
\label{eq:sumlmk}
\Sigma \, T = \sum_{\ell mk} T_{\ell mk}(r)\, S_{\ell mk}(\theta)\, e^{-i\omega_{mk} t+im\vp} ,
\end{align}
with a similar form for $\Phi$.  The separated field equations \eqref{eq:spheroidH} and \eqref{eq:radTeukEq} have the same form except all the continuous-in-$\omega$ amplitudes are replaced\footnote{We use both notations in this paper, depending on the context. For example, Sec.~\ref{sec:expanding} on expanding the scalar field is not limited to a discrete frequency, so we use $\ell m \omega$ in that section.} with discrete ones: $R_{\ell m\omega}\to R_{\ell mk}$, $T_{\ell m\omega}\to T_{\ell mk}$, $S_{\ell m}(\theta;-a^2 \omega^2) \to S_{\ell mk}(\theta)$, and $\lambda_{\ell m\omega} \to \lambda_{\ell mk}$. Orthogonality of the spheroidal harmonics is then used to pluck off the source amplitudes
\begin{align}
    T_{\ell mk} = \frac{q}{T_\theta}\int_{0}^{T_\theta} \frac{S_{\ell mk}(\theta_p(t))}{u^t(\theta_p(t))}e^{i\omega_{mk} t-im\vp_p(t)}\delta(r-r_0)dt , 
    \label{eq:sourcemodes}
\end{align}
with $T_{\ell mk}(r) = T^{(0)}_{\ell mk} \, \delta(r - r_0)$ and $T_\theta$ being the period of polar libration.

Solutions of the radial equation \eqref{eq:radTeukEq} are then sought that have the asymptotic behavior \cite{TeukPres74}
\begin{align}
\label{eq:Rinf}
R_{\ell mk}(r) &= Z^{+}_{\ell mk} r^{-1} e^{i \omega_{mk} r_*} , \qquad r \to \infty ,  \\
\label{eq:Rhor}
R_{\ell mk}(r) &= Z^{-}_{\ell mk} e^{-i \gamma_{mk} r_*} , \quad \qquad r \to r_{+} ,
\end{align}
where the wavenumber $\gamma = \omega - m a/(2 M r_{+})$ is also evaluated at the discrete frequencies, and where $r_{\pm} = M\pm\sqrt{M^2-a^2}$ is the outer horizon radius and $r_*$ is the tortoise coordinate, related to $r$ by 
\begin{align}
    r_* = r + \frac{2M r_+}{r_+ - r_-} \log \frac{r-r_+}{2M} -  \frac{2M r_-}{r_+ - r_-} \log \frac{r-r_-}{2M}.
\end{align}
The primary goal is to determine the amplitudes $Z^{+}_{\ell mk}$ and $Z^{-}_{\ell mk}$.  To do so, we use two independent, unit-normalized solutions $R^{-}_{\ell mk}$ and $R^{+}_{\ell mk}$ of the homogeneous equation that are a scaled version of the ``in-up'' basis, $R^{\text{in}}_{\ell mk}$ and $R^{\text{up}}_{\ell mk}$ \cite{SasaTago03}.  

Using these basis solutions, we then construct a Green function
\begin{align}
    G_{\ell mk}(r,r') = \frac{R_{\ell mk}^{+}(r_>)R_{\ell mk}^{-}(r_<)}{W_{\ell mk}}, \label{eq:greenfunc}
\end{align}
where the (constant) Wronskian is 
\begin{align}
    W_{\ell mk} = \Delta\left(R_{\ell mk}^{-}\frac{d}{dr}R_{\ell mk}^{+}-R_{\ell mk}^{+}\frac{d}{dr}R_{\ell mk}^{-}\right),
\end{align}
and where $r_> = \text{max}(r,r')$ and $r_< = \text{min}(r,r')$. The solution of the full inhomogeneous radial equation is then given by 
\begin{align}
\label{eq:intsolution}
    R_{\ell mk}(r) = -4\pi\int_{r_+}^{\infty} G_{\ell mk}(r,r')T_{\ell mk}(r')dr'.
\end{align}

As Eqs. \eqref{eq:sourcemodes} makes clear, the source in this application is localized to the sphere $r = r_0$, so the solution to \eqref{eq:intsolution} is
\begin{align}
    R_{\ell mk}(r) = &-\frac{4\pi T^{(0)}_{\ell mk}}{W_{\ell mk}}\bigg(R_{\ell mk}^{-}(r_0)R_{\ell mk}^{+}(r)\Theta(r-r_0) \notag \\
    &+ R_{\ell mk}^{+}(r_0)R_{\ell mk}^{-}(r)\Theta(r_0-r)\bigg) ,
\end{align}
where $\Theta(x)$ is the Heaviside function.  From this solution we can read off immediately the asymptotic amplitudes defined in \eqref{eq:Rinf} and \eqref{eq:Rhor}, finding 
\begin{align}
    Z^{\pm}_{\ell mk} = -\frac{4\pi q R^{\mp}_{\ell mk}(r_0)}{T_\theta W_{\ell mk}} 
    \int_{0}^{T_\theta}e^{i\omega t-im\vp_p(t)} \frac{S_{\ell mk}(\theta_p(t))}{u^t}dt. \label{eq:asympAmp}
\end{align}
Note that other normalizations of the two basis functions can be, and often are, used (including the standard definition of $R^{\text{in}}_{\ell mk}$ and $R^{\text{up}}_{\ell mk}$ \cite{SasaTago03}), in which case an added multiplicative factor appears in \eqref{eq:asympAmp}.

Once these amplitudes are known, the complex asymptotic solutions \eqref{eq:Rinf} and \eqref{eq:Rhor} for $R_{\ell mk}$ then contribute to the construction of the stress-energy tensor for the scalar field, and whence on Kerr the expression of conserved energy and angular momentum currents, $-{T^{\mu}}_{\nu} \xi^{\nu}_{(t)}$ and ${T^{\mu}}_{\nu} \xi^{\nu}_{(\vp)}$, respectively.  These in turn are projected on surface normals of bounding spheres near infinity and the horizon and integrated \cite{TeukPres74} to yield the total time-averaged fluxes
\begin{align}
    \left\langle\frac{dE}{dt}\right\rangle_{\infty} &= \sum_{\ell mk} \frac{\omega_{mk}^2}{4\pi}\left|Z_{\ell mk}^+\right|^2, \\
    \left\langle\frac{dE}{dt}\right\rangle_{\mathcal{H}} &= \sum_{\ell mk} \frac{M r_+\omega_{mk}}{2\pi}\left(\omega_{mk} - \frac{ma}{2Mr_+}\right) \left|Z_{\ell mk}^{-}\right|^2, \\
    \left\langle\frac{dL_z}{dt}\right\rangle_{\infty} &= \sum_{\ell mk} \frac{m\omega_{mk}}{4\pi}\left|Z_{\ell mk}^{+}\right|^2, \\
    \left\langle\frac{dL_z}{dt}\right\rangle_{\mathcal{H}} &= \sum_{\ell mk} \frac{M r_+ m}{2\pi}\left(\omega_{mk}-\frac{ma}{2Mr_+}\right)\left|Z_{\ell mk}^{-}\right|^2.
\end{align}

\subsection{Expanding the Scalar Field}
\label{sec:expanding}

We now set about finding analytic solutions for the radial homogeneous solutions (basis functions) and the angular functions.  The beginning of the process is the same as what is done for numerical calculations, and readily allows analytic PN expansion. 

Throughout this subsection, we return to notation where $\omega$ is in principle not a discrete multiple of the fundamental frequency, since the methods are more broadly applicable demanding only that the frequency is small and scales appropriately with respect to the radial field point.

\subsubsection{Angular functions}

First, we work out the PN expansion of the spheroidal harmonic functions, which are solutions of Eq.~\eqref{eq:spheroidH}.  Part of this calculation precedes the expansion of the radial mode functions, as we must first determine the separation eigenvalues $\lambda_{\ell m\omega}$. Our previous paper \cite{CastETC25a} discussed a procedure to PN expand generic spin-weighted spheroidal Legendre functions of $\th$ for arbitrary $\ell$, $m$, and $\omega$, where the spheroidicity factor $a \omega$ is a small parameter in the PN limit \cite{GanzETC07,SagoFuji15,FujiShib20,IsoyETC22}.  

In our present application, $s=0$ and we specialize our summary of the procedure for this particular case.  The first step is to PN expand the azimuthal function $\varphi(\chi)$.  However, $\varphi(\chi)$ enters into the calculation of the source amplitudes in Eq. \eqref{eq:asympAmp} only through the factor $e^{im\varphi}$.  When we PN expand this term, the part that survives in the Newtonian limit has an important particular form that can be factored out.  We find \cite{Cast25}
\begin{align}
    e^{im\varphi} = \left(\frac{x\cos\chi\pm i\sin\chi}{\sin\theta}\right)^{|m|}\left[1+\frac{2ia\chi}{p^{3/2}}+O\left(\frac{1}{p^{2}}\right)\right] , 
\end{align}
where the sign inside the first term is determined by the sign of $m$.  What is important to note is that the $(\sin\theta)^{-|m|}$ factor will cancel out a corresponding factor in the functional dependence of the spheroidal Legendre function of the same order in $m$.

To calculate the full $s=0$ spheroidal harmonics, we use the approach developed and used many times previously (e.g., \cite{PresTeuk73,Hugh00b,CasaOtte05,Dola07,WarbBara10,NasiOsbuEvan19}) of expressing these functions as series in ordinary spherical harmonics, $Y_{lm}(\theta,\varphi)$
\begin{align}
    S_{\ell m\omega}(\theta)e^{im\vp} = \sum_{j=0}^{\infty}b_{jm}^{\ell}(a\omega)Y_{jm}(\theta,\varphi) . \label{eq:spheroiH}
\end{align}
Here the coefficients $b_{jm}^{\ell}(a\omega)$ depend on the orders, $\ell$ and $m$, of the target harmonic but range over a potentially infinite set of $j$.  In practice, the coefficients are known to fall off exponentially with the separation $|\ell - j|$ and to vanish (when $s=0$) for odd values of $\ell -j$.  This property is nicely demonstrated in Fig.~1 of \cite{WarbBara10}.  

This behavior is explained by making a series expansion of the coefficients in powers of $a\omega$, which is of course a PN expansion.  The leading behavior of the coefficients is found to be $b_{jm}^{\ell}(a\omega)\sim (a\omega)^{|\ell - j|}$, allowing us to express the spheroidal harmonics as a double series  
\begin{align}
    S_{\ell m\omega}(\theta)e^{im\vp}=\sum_{n=0}^{\infty}(a\omega)^{n}\left[\sum_{i=-n}^{n} \, d_{in}^{\ell m}Y_{\ell+i,m}(\theta,\varphi)\right].
\end{align}
For the $s=0$ case, the series expansion ranges over only even powers of $a \omega$ \cite{PresTeuk73}.
The coefficients $d_{in}^{\ell m}$ are readily available in the \texttt{SpinWeightedSpheroidalHarmonics} package of the Black Hole Perturbation Toolkit \cite{SpinWeightedSpheroidalHarmonics,BHPTK18}.

Proceeding further, the spherical harmonics have dependence on $\theta$ of the form 
\begin{align}
    Y_{\ell m}(\theta,0) = (\sin\theta)^{|m|}\sum_{n=0}^{\ell-|m|}c_n(\cos\theta)^n,
\end{align}
where $c_n$ are constants.  As reference to Eq.~\eqref{eq:spheroiH} shows, this structure maps through to the spheroidal harmonics.  Converting from $\th$ to $\chi$ using $\cos\theta = \sqrt{1-x^2}\cos\chi$ and combining with the previous steps, we obtain the following schematic structure for the PN expansion of the spheroidal harmonics 
\begin{align}
    S_{\ell m\omega}&(\theta)e^{im\vp}=\left(x\cos\chi \pm i\sin\chi\right)^{|m|}\times \notag\\ 
    &\left(\text{polynomial in }\chi\right)\times\left(\text{polynomial in }\cos\chi\right) .
\end{align}
The degree of the polynomial in $\cos\chi$ is dependent on $\ell$, $m$, and the PN order (via powers of $a\omega$) and the degree of the polynomial in $\chi$ is dependent on the PN order of the expansion of the phase.

\subsubsection{Teukolsky radial functions}

Because the radial equation Eq. \eqref{eq:radTeukEq} is the Teukolsky $s=0$ equation, the methods pioneered by Mano, Suzuki and Takasugi (MST) \cite{ManoSuzuTaka96b} can be utilized.  In the MST method, the homogeneous solutions to Eq. \eqref{eq:radTeukEq} are represented as a convergent sum of hypergeometric functions (see also \cite{SasaTago03}).  In this paper we use the Coulomb wave function representation, which we briefly review below (see sections 4.3 and 4.4 of \cite{SasaTago03}).  For another detailed discussion, see Appendix B of \cite{Thro10}.

\begin{widetext}
The Coulomb wave function representation begins with defining a particular set of solutions $R_C$ (with mode indices suppressed) to the Teukolsky equation for arbitrary $s$ \cite{Leav86,SasaTago03} as an infinite series
\begin{align}
\label{eq:MSTC}
    R_{C}^{\nu} =& \left(1-\frac{\epsilon\kappa}{z}\right)^{-s-i(\epsilon+\tau)/2}e^{-iz}2^{s}(2z)^{\nu-s}\frac{\Gamma(\nu+1-s+i\epsilon)}{\Gamma(\nu+1+s-i\epsilon)}\times \\
    &\sum_{j=-\infty}^{\infty}a_j^{\nu}(2z)^{j}\frac{\Gamma(j+\nu+1+s-i\epsilon)}{\Gamma(2j+2\nu+2)}M(j+\nu+1-s+i\epsilon,2j+2\nu+2;2iz), \notag
\end{align}
where $\epsilon = 2M\omega$, $\epsilon_+ = (\epsilon+\tau)/2$, $\kappa = \sqrt{1-\tilde{a}^2}$, $\tau = (\epsilon-m\tilde{a})/\kappa$, and $\tilde{a} = a/M$.  The functions $M(a,b;z)$ are the regular confluent hypergeometric functions whose argument $z = (r-M(1-\kappa))\omega$ is a scaling of the radial coordinate.  These solutions can conversely be thought of as infinite series over the closely related Coulomb wave functions.  Coulomb wave functions have a long history of use in similar problems in physics \cite{YostWheeBrei36}.   The parameter $\nu$ is the renormalized angular momentum whose eigenvalues allow the doubly infinite series to converge as $j \to \pm \infty$.  Along with $R_C^{-\nu-1}$, which also converges, we have two independent homogeneous solutions to the Teukolsky radial equation.

With $R_C^{\nu}$ and $R_C^{-\nu-1}$, two linear combinations can be formed for radial functions that share the desired asymptotic behavior of $R^{-}_{\ell m \omega}$ (at the horizon) and $R^{+}_{\ell m \omega}$ (at infinity) \cite{SasaTago03,Thro10}.  These are, respectively,
\begin{align}
    {}_sR_{\ell m\omega}^{\text{in}} &= R_C^\nu + \frac{K_{-\nu-1}}{K_\nu} R_C^{-\nu-1}, \label{eq:MSTin}\\
    {}_sR_{\ell m\omega}^\text{{up}} &= e^{-\pi\epsilon-i\pi s}\frac{\sin(\pi(\nu+s-i\epsilon))}{i\sin(2\pi\nu)}\left[R_C^{-\nu-1}+ie^{-i\pi\nu}\frac{\sin(\pi(\nu-s+i\epsilon))}{\sin(\pi(\nu+s-i\epsilon))}R_{C}^{\nu}\right], \label{eq:MSTup}
\end{align}
where the factor $K_\nu$ depends on $\omega$ but not on the radial coordinate, see Eq.~(165) of \cite{SasaTago03}. As already discussed in our earlier paper \cite{CastETC25a} (see footnote 2) Eq.~\eqref{eq:MSTin} differs from Eq.~(166) of \cite{SasaTago03} by dividing a normalization factor of $K_\nu$. We find this significantly easier to PN expand to high order for two reasons. The first is that, while $K_\nu$ and $K_{-\nu-1}$ are both quite lengthy, their ratio benefits from a number of cancellations. This ratio is also known as \textit{tidal response function} \cite{BautETC24}. The second reason is that for most cases the ratio scales as $\omega^{2\ell-1}$ which suppresses $R_C^{-\nu-1}$ with respect to $R_C^\nu$. 
\end{widetext}

With the MST method we construct high-order PN expansions of the Teukolsky mode functions in three steps, similar to the order taken in a numerical calculation.  First, we find a PN expansion of the renormalized 
angular momentum, $\nu = \nu(s,l,m,\omega)$, by terminating a continued fraction calculation at a high enough depth, which is related to desired PN order.  Second, using $\nu$, we construct PN expansions of the series coefficients, $a_{j}^{\nu}$, which involves PN expanding the gamma function factors.  Then finally, we assemble the PN expansion of the sum in Eq.~\eqref{eq:MSTC} and then the mode functions in Eqs. \eqref{eq:MSTin} and \eqref{eq:MSTup}.  More detail on the procedure can be found in \cite{KavaOtteWard15,KavaOtteWard16}.

\subsubsection{Sum over modes}

The sum over modes in Eq.~\eqref{eq:sumlmk} is formally infinite in both $\ell$ and $k$.  In practice, however, for a given PN order, both sums can be truncated to a finite number of terms.  For the sum in $\ell$, the leading PN order behavior of the radial functions is found to be $\ell$-dependent \cite{KavaOtteWard15,KavaOtteWard16,Munn20}, with a target PN order $N$ requiring only a partial sum up to some $\ell_\text{max}$.\footnote{This is true in the dissipative case.  For the conservative part of the self-force, the full sum over $\ell$ is needed, and a separate general-$\ell$ ansatz is then used to solve the radial equation.}  For the sum over $k$, because of the strong falloff in the angular harmonics with frequency, we find the $k$ modes for a given $\ell$, $m$ and PN order $N$ to lie in the following range
\begin{align}
     -l-m-2\lfloor N/4\rfloor\leq k \leq l-m+2\lfloor N/4 \rfloor,
\end{align}
but only for those $k$ such that $\ell+m+k$ is even.  This range of $k$ is the same as in the gravitational case \cite{CastETC25a}, but with the extra restriction on $\ell+m+k$ being even.     

\begin{widetext}

\section{Results}

\subsection{Structure}

The general structure of the PN expansion of the scalar energy flux can be written similar to that in the gravitational case \cite{Blan14,CastETC25a}, with only a difference in the overall Newtonian-limit prefactor.  The infinity-side scalar energy flux has the following general expression
\begin{align}
    \begin{autobreak}
    \left\langle \frac{dE}{dt} \right\rangle_{\infty} =
    \frac{1}{3}q^2p^{-4}\bigg[\mathcal{A}_0 
    + \mathcal{A}_1 p^{-1} 
    + \mathcal{A}_{3/2} p^{-3/2} 
    + \mathcal{A}_{2} p^{-2} 
    + \mathcal{A}_{5/2}p^{-5/2} 
    + \left(\mathcal{A}_{3} + \mathcal{A}_{3L}\log(p)\right)p^{-3}
    + \mathcal{A}_{7/2}p^{-7/2}
    + \left(\mathcal{A}_{4}+\mathcal{A}_{4L}\log(p)\right)p^{-4}
    + \left(\mathcal{A}_{9/2}+\mathcal{A}_{9/2L}\log(p)\right)p^{-9/2}
    + \left(\mathcal{A}_{5}+\mathcal{A}_{5L}\log(p)\right)p^{-5}
    + \left(\mathcal{A}_{11/2}+\mathcal{A}_{11/2L}\log(p)\right)p^{-11/2}
    + \left(\mathcal{A}_{6}+\mathcal{A}_{6L}\log(p) + \mathcal{A}_{6L2}\log^{2}(p)\right)p^{-6} 
    + \cdots\bigg],
    \end{autobreak}
\end{align}
with each coefficient $\mathcal{A}_{iLj}$ representing a function of $a$ and $x$. We use the same coefficient conventions in \cite{CastETC25a}, where the tag $i$ represents the integer, or half-integer, relative PN order of the term and the tag $Lj$ also appears when a power-of-log term, $\log^{j}(p)$, is present. We also use the following abbreviations $\mathcal{A}_{iL0} = A_{i}$ and $\mathcal{A}_{iL1} = \mathcal{A}_{iL}$ for simplicity. The general expression for the angular momentum fluxes at infinity has a similar form
\begin{align}
    \begin{autobreak}
        \left\langle \frac{dL_z}{dt} \right\rangle_{\infty} = 
            \frac{1}{3}q^2 x p^{-5/2}\bigg[\mathcal{C}_0
            +\mathcal{C}_1 p^{-1}
            +\mathcal{C}_{3/2} p^{-3/2}
            +\mathcal{C}_{2} p^{-2}
            +\mathcal{C}_{5/2} p^{-5/2}
            +\left(\mathcal{C}_{3}+\mathcal{C}_{3L}\log(p)\right)p^{-3}
            +\mathcal{C}_{7/2}p^{-7/2}
            +\left(\mathcal{C}_{4}+\mathcal{C}_{4L}\log(p)\right)p^{-4}
            +\left(\mathcal{C}_{9/2}+\mathcal{C}_{9/2L}\log(p)\right)p^{-9/2}
            +\left(\mathcal{C}_{5}+\mathcal{C}_{5L}\log(p)\right)p^{-5}
            +\left(\mathcal{C}_{11/2}+\mathcal{C}_{11/2L}\log(p)\right)p^{-11/2}
            +\left(\mathcal{C}_{6}+\mathcal{C}_{6L}\log(p)+\mathcal{C}_{6L2}\log^{2}(p)\right)p^{-6}
            +\cdots\bigg].
    \end{autobreak}
\end{align}
We note that there is an overall prefactor of $x$ in the angular momentum flux. The angular momentum $\mathcal{L}_z$ is defined along a specific axis of rotation, thus, in the nonspinning limit, $x \neq 1$ implies an orbital plane not perpendicular to the specified axis of rotation. 
\end{widetext}

In what follows, we present the full scalar energy and angular momentum flux expressions up to 3.5PN to examine the low-order behavior of the spin and inclination terms and display only a subset of higher-order terms where we focus entirely on the spin and inclination-dependent aspects.  Because the flux expressions become increasingly unwieldy with increasing PN order, we relegate the full 12PN expressions for the scalar fluxes to online repositories \cite{BHPTK18,PostNewtonianSelfForce}.

To facilitate this examination, the flux components $\mathcal{A}_{iLj}$ and $\mathcal{C}_{iLj}$ are further broken down into 
\begin{align}
    \mathcal{A}_{iLj}(a,x) &= \mathcal{A}_{iLj}^{(0)} + \sum_{k=0}\mathcal{A}_{iLj}^{Sk}(a,x), \label{eq:ECBD} \\
    \mathcal{C}_{iLj}(a,x) &= \mathcal{C}_{iLj}^{(0)} + \sum_{k=0}\mathcal{C}_{iLj}^{Sk}(a,x), \label{eq:AMCBD}
\end{align}
where $\mathcal{A}_{iLj}^{(0)}$ and $\mathcal{C}_{iLj}^{(0)}$ are the non-spinning limit of $\mathcal{A}_{iLj}$ and $\mathcal{C}_{iLj}$ respectively, while $\mathcal{A}_{iLj}^{Sk}$ and $\mathcal{C}_{iLj}^{Sk}$ are components proportional to $a^k$. This decomposition is chosen to separate the non-spinning components from the spin contributions, so that all of the $\mathcal{A}_{iLj}^{Sk}$ and $\mathcal{C}_{iLj}^{Sk}$ vanish as $a \to 0$.

\subsection{Scalar Energy Flux at Infinity}

The complete scalar energy flux expression up to 3.5PN relative order is
\begin{align}
    \begin{autobreak}
    \MoveEqLeft
    \left\langle \frac{dE}{dt} \right\rangle_{\infty} =
        \frac{1}{3}q^2 p^{-4} \bigg[ 1
            - 2 p^{-1}
            + (2\pi 
            -4\tilde{a}x)p^{-3/2}
            + \bigg( -10
            -\frac{5\tilde{a}^2}{2}
            +\frac{7\tilde{a}^2x^2}{2}\bigg)p^{-2}
            + \bigg(\frac{12\pi}{5}
            +4\tilde{a}x\bigg)p^{-5/2}
            + \bigg(\frac{1331}{75} 
            +9\tilde{a}^2
            -\frac{76}{15}\gamma
            +\frac{38}{15}\log(p)
            +\frac{4}{3}\pi^2
            -10\tilde{a}\pi x
            +3\tilde{a}^2x^2
            -\frac{76}{15}\log(2)\bigg)p^{-3}
            +\bigg(-\frac{521\pi}{14}
            -\frac{13\tilde{a}^2\pi}{2}
            +\frac{278\tilde{a}x}{5}
            +8\tilde{a}^3x
            +\frac{17}{2}\tilde{a}^2\pi x^2
            -12\tilde{a}^3x^3\bigg)p^{-7/2}
        +O(p^{-4})\bigg].
    \end{autobreak} \label{eq:energyfluxI}
\end{align}
Several features are immediately evident.  First, the expression above in the $\tilde{a} \to 0$ limit matches the scalar flux for circular orbits about a Reissner-N\"{o}rdstrom black hole \cite{BiniCarvGera16} if we set the electric charge on the hole to zero.  Second, $x$ is multiplied with $\tilde{a}$ in a manner where the sign difference between them defines the orientation of the orbit.  That is, $\mathrm{sgn}(\tilde{a}x) = 1$ for prograde orbits, while $\mathrm{sgn}(\tilde{a}x) = -1$ for retrograde orbits.  Third, at each order in the expansion, $x$ enters as a finite polynomial, suggesting a simple functional dependence on inclination.

\setlength\extrarowheight{3pt}
\begin{table*}[t]
    \centering
    \caption{List of higher-order components of the scalar energy flux at infinity from 4PN up to 5 PN and a select 6PN flux component.}
    \begin{tabular}{|c||c|}
        \hline 
        Flux Component & Flux Expression \\[3pt]
        \hline
        $\mathcal{A}_{4}^{(0)}$ & $\frac{152 \pi ^2}{15}-\frac{6392 \gamma
    }{525}+\frac{1200581}{6125} -\frac{8872 \log (2)}{175}$ \\[3pt] 
        \hline
        $\mathcal{A}_{4L1}^{(0)}$ & $\frac{3196}{525}$ \\[3pt]
        \hline
        $\mathcal{A}_{4}^{S1}$ & $-\frac{164}{5}\tilde{a}\pi x$ \\[3pt]
        \hline
        $\mathcal{A}_{4}^{S2}$ & $\tilde{a}^2\left(\frac{25}{2}-\frac{55}{2}x^2\right)$ \\[3pt]
        \hline
        $\mathcal{A}_{4}^{S4}$ & $\tilde{a}^4\left(\frac{87}{16}-\frac{99}{8}x^2+\frac{111}{16}x^4\right)$ \\[3pt] \hline 
        $\mathcal{A}_{9/2}^{(0)}$ & $-\frac{152 \gamma  \pi }{15}-\frac{42017 \pi
    }{675}-\frac{152}{15} \pi  \log 2$ \\[3pt] \hline 
        $\mathcal{A}_{9/2L1}^{(0)}$ & $\frac{76}{15}\pi$ \\[3pt]
        \hline
        $\mathcal{A}_{9/2}^{S1}$ & $\frac{842 \tilde{a} x}{175}-8 \pi ^2 \tilde{a} x+\frac{184 \gamma  \tilde{a}x}{5}+\frac{184}{5} \tilde{a} x \log 2 + \frac{8}{3}\tilde{a}x\log\kappa + \frac{4}{3}\tilde{a}x\Psi^{(0,1)}(\tilde{a})$ \\[3pt] \hline
        $\mathcal{A}_{9/2L1}^{S1}$ & $-\frac{296}{15}\tilde{a}x$ \\[3pt] \hline
        $\mathcal{A}_{9/2}^{S2}$ & $\tilde{a}^2\pi\left(\frac{29}{5}+41x^2\right)$ \\[3pt] \hline
        $\mathcal{A}_{9/2}^{S3}$ & $\tilde{a}^3\left(\frac{40}{3}x-48x^3\right)$ \\[3pt] \hline
        $\mathcal{A}_{5}^{(0)}$ & $-\frac{404 \pi ^2}{15}+\frac{1193924 \gamma
    }{11025} -\frac{779720203}{3472875} -\frac{369603 \log 3}{2450}+\frac{881908 \log 2}{2205}$ \\[3pt]
    \hline
    $\mathcal{A}_{5L1}^{(0)}$ & $-\frac{596962}{11025}$ \\[3pt]
    \hline
    $\mathcal{A}_{5}^{S1}$ & $\frac{12861}{70}\tilde{a}\pi x$ \\[3pt]
    \hline
    $\mathcal{A}_{5}^{S2}$ & $\tilde{a}^2\left(\frac{20 \pi ^2 x^2}{3}-\frac{76 \gamma  x^2}{3}-\frac{1496 x^2}{15}-\frac{76}{3} x^2 \log 2 -\frac{16 \pi^2}{3}+\frac{304 \gamma }{15}-\frac{2803}{25}+\frac{304 \log 2}{15}\right)$ \\[3pt]
    \hline
    $\mathcal{A}_{5L1}^{S2}$ & $\tilde{a}^2\left(-\frac{152}{15}+\frac{38}{3}x^2\right)$ \\[3pt]
    \hline
    $\mathcal{A}_{5}^{S3}$ & $\tilde{a}^3\pi\left(\frac{25}{2}x-\frac{45}{2}x^3\right)$, \\[3pt]
    \hline
    $\mathcal{A}_{5}^{S4}$ & $\tilde{a}^4\left(-\frac{261}{8}+\frac{69}{4}x^2+\frac{219}{8}x^4\right)$ \\[3pt] 
    \hline
    $\mathcal{A}_{6}^{S0}$ & $-\frac{4}{3}(1+x^2)\left(\Psi^{(0,1)}(\tilde{a})+2\gamma\right)+\frac{32}{15}(1-\kappa)-\frac{16}{3}\log\kappa$ \\[3pt] 
    \hline
    \end{tabular}
    \label{tab:EFluxComp}
\end{table*}

Higher-order (spin-dependent) components of the energy flux up to 5PN are presented in Table \ref{tab:EFluxComp}.  At higher PN order, non-polynomial functions of $a$ begin to appear.  Notable among these are real functions $\Psi^{(r,\sigma)}(\tilde{a})$ and $\bar{\Psi}^{(r,\sigma)}(\tilde{a})$ 
that emerge as combinations of the polygamma functions $\psi^{(r)}$
\begin{align}
    \Psi^{(r,\sigma)}(\tilde{a}) &= \psi^{(r)}\left(1+\frac{i\sigma \tilde{a}}{\kappa}\right) + \psi^{(r)}\left(1-\frac{i\sigma\tilde{a}}{\kappa}\right),  \notag \\
    i\bar{\Psi}^{(r,\sigma)}(\tilde{a}) &= \psi^{(r)}\left(1+\frac{i\sigma\tilde{a}}{\kappa}\right)-\psi^{(r)}\left(1-\frac{i\sigma\tilde{a}}{\kappa}\right). \label{eq:polygamma}
\end{align}
Here the complex argument depends on $\tilde{a}$ and an integer $\sigma$. Starting with $\mathcal{L}_{9/2}^{S1}$ onward, these polygamma functions become increasingly common at higher PN order. In particular, we note $\mathcal{L}_{6}^{S0}$ that has no leading power of $a$, and whose sole dependence on $a$ comes from non-polynomial functions.  The presence of non-polynomial functions of $a$ in the higher-order components of the flux was previously seen in the gravitational case for both dissipative \cite{TaraETC13,Shah14,Fuji15,CastETC25a,SagoFujiNaka25} and conservative \cite{KavaOtteWard16} quantities.

\subsection{Scalar Angular Momentum Flux at Infinity}

The full angular momentum flux expression up to 3.5PN relative order is
\begin{align}
    \begin{autobreak}
        \MoveEqLeft
        \left\langle\frac{dL_z}{dt}\right\rangle = \frac{1}{3}q^2 x p^{-5/2} \bigg[ 
            1 - 2p^{-1}
            +\bigg(2\pi
            +\frac{3\tilde{a}}{x}
            -6\tilde{a}x\bigg)p^{-3/2}
            +\left(-10-\frac{7}{2}\tilde{a}^2+\frac{9}{2}\tilde{a}^2x^2\right)p^{-2}
            +\bigg( \frac{12\pi}{5}
            -\frac{10\tilde{a}}{x}
            +12\tilde{a}x\bigg)p^{-5/2}
            +\bigg(\frac{1331}{75}
            -\frac{5\tilde{a}^2}{2}
            -\frac{76}{15}\gamma
            +\frac{38}{15}\log(p)
            +\frac{4\pi^2}{3}
            +\frac{8\tilde{a}\pi}{x}
            -16\tilde{a}\pi x
            +\frac{21\tilde{a}^2x^2}{2}
            +\frac{76}{15}\log(2)\bigg)p^{-3}
            +\bigg(-\frac{512\pi}{14}
            -10\tilde{a}^2\pi
            -\frac{129\tilde{a}}{5x}
            -\frac{55\tilde{a}^3}{8x}
            +\frac{357\tilde{a}x}{5}
            +\frac{127\tilde{a}^3x}{4}
            +12\tilde{a}^2x^2
            -\frac{223\tilde{a}^3x^3}{8}
            \bigg)p^{-7/2}
        +O(p^{-4})\bigg].
    \end{autobreak}
\end{align}
Aside from the previously mentioned prefactor in the angular momentum case, the same observations as noted about the energy flux hold for this expression.  With $x$ present in the prefactor, most of the terms in the angular momentum flux vanish as $x \to 0$.  The few surviving terms give rise to a 1.5PN weaker flux in the polar-orbit $(x\to 0)$ limit  
\begin{align} 
    \begin{autobreak}
    \MoveEqLeft
    \lim_{x \to 0} \left\langle\frac{dL_z}{dt}\right\rangle_{\infty} = \frac{1}{3}q^2\tilde{a} p^{-4} \bigg[
        3 - 10p^{-1}
        + 8\pi p^{-3/2}
        + \left(-\frac{129}{5}-\frac{55}{8}\tilde{a}^2\right)p^{-2}
    +O(p^{-5/2})\bigg].
    \end{autobreak} \label{eq:angfluxPolar}
\end{align}
This polar-orbit angular momentum flux vanishes completely in the $a = 0$ limit, making this a frame dragging effect as the rotation of the hole causes the plane of the polar orbit to precess.   

Higher-order components of the scalar angular momentum flux at infinity are presented in Table \ref{tab:LFluxComp}, complete up to 5PN. The full scalar angular momentum flux at infinity through 12PN can be obtained from the online repositories \cite{BHPTK18,PostNewtonianSelfForce,UNCGrav22}.

\begin{table*}[t]
    \centering
    \caption{List of higher-order components of the scalar angular momentum flux from 4PN up to 5PN.}
    \begin{tabular}{| c || c |}
       \hline
        Flux Component & Flux Expression \\[3pt]
        \hline
        $\mathcal{C}_{4}^{(0)}$ & $\frac{152 \pi ^2}{15}-\frac{6392 \gamma}{525}+\frac{1200581}{6125}-\frac{8872 \log 2}{175}$ \\[3pt]
        \hline
        $\mathcal{C}_{4L}^{(0)}$ & $\frac{3196}{525}$ \\[3pt] 
        \hline
        $\mathcal{C}_{4}^{S1}$ & $-\frac{112 \tilde{a}\pi  x}{5}-\frac{8 a\pi }{x}$ \\[3pt] 
        \hline
        $\mathcal{C}_{4}^{S2}$ & $\tilde{a}^2(82-93 x^2)$ \\[3pt] 
        \hline
        $\mathcal{C}_{4}^{S4}$ & $\tilde{a}^4\left(\frac{83 x^4}{8}-\frac{65 x^2}{4}+\frac{47}{8}\right)$ \\[3pt] 
        \hline 
        $\mathcal{C}_{9/2}^{(0)}$ & $-\frac{152 \gamma  \pi }{15}-\frac{42017 \pi}{675}-\frac{152}{15} \pi  \log 2$ \\[3pt] 
        \hline
        $\mathcal{C}_{9/2L}^{(0)}$ & $\frac{76 \pi  }{15}$ \\[3pt]
        \hline
        $\mathcal{C}_{9/2}^{S1}$ & $\frac{\tilde{a}}{x}\left[\frac{2}{3}\left(1+x^2\right)\Psi^{(0,1)}(a)+\frac{4}{3} \left(1+x^2\right) \log \kappa-\frac{20 \pi^2}{3}\left(2x^2-1\right)+\frac{4}{15}\left(\gamma+\log2\right)\left(202x^2-83\right)-\frac{18092 x^2}{175}+\frac{66119}{525}\right]$ \\[3pt]
        \hline
        $\mathcal{C}_{9/2L}^{S1}$ & $-\frac{138 \tilde{a}x}{5}+\frac{52 a}{5 x}$ \\[3pt]
        \hline
        $\mathcal{C}_{9/2}^{S2}$ & $83 \pi \tilde{a}^2 x^2-\frac{231 \tilde{a}^2\pi }{5}$ \\[3pt]
        \hline
        $\mathcal{C}_{9/2}^{S3}$ & $\tilde{a}^3\left(-10 x^3-\frac{197 x}{6}+\frac{121}{6 x}\right)$ \\[3pt]
        \hline
        $\mathcal{C}_{5}^{(0)}$ & $-\frac{404 \pi ^2}{15}+\frac{1193924 \gamma}{11025}-\frac{779720203}{3472875}-\frac{369603 \log 3}{2450}+\frac{881908 \log2}{2205}$ \\[3pt] 
        \hline
        $\mathcal{C}_{5L}^{(0)}$ & $-\frac{596962}{11025}$ \\[3pt]
        \hline
        $\mathcal{C}_{5}^{S1}$ & $\tilde{a}\left(\frac{1414 \pi  x}{5}-\frac{954 \pi }{7 x}\right)$ \\[3pt] 
        \hline
        $\mathcal{C}_{5}^{S2}$ & $\tilde{a}^2\bigg(10 \pi ^2 x^2-38 \gamma  x^2-\frac{5077 x^2}{30}-38 x^2 \log 2 -\frac{26 \pi ^2}{3}+\frac{494 \gamma}{15}+\frac{649}{50}+\frac{494 \log 2}{15}\bigg)$ \\[3pt]
        \hline
        $\mathcal{C}_{5L}^{S2}$ & $\tilde{a}^2\left(19 x^2-\frac{247}{15}\right)$ \\[3pt]
        \hline
        $\mathcal{C}_{5}^{S3}$ & $\tilde{a}^3 \left(-\frac{185 \pi  x^3}{2}+109 \pi  x-\frac{49 \pi }{2 x}\right)$ \\[3pt] 
        \hline
        $\mathcal{C}_{5}^{S4}$ & $\tilde{a}^4\left(\frac{1535 x^4}{16}-\frac{935 x^2}{8}+\frac{463}{16}\right)$ \\[3pt]
        \hline
    \end{tabular}
    \label{tab:LFluxComp}
\end{table*}

\subsection{Leading-spin terms}

Our previous paper on gravitational fluxes \cite{CastETC25a} singled out the leading-spin contributions --- flux terms that hold the first appearance of a new higher power of $a$.  Similar sequences exist in the scalar fluxes. Here we find the leading-spin terms through 8PN in the energy flux are
\begin{align}
    \mathcal{A}_{3/2}^{S1} =& -4\tilde{a}x, \\
    \mathcal{A}_{2}^{S2} =& -\frac{1}{2}\tilde{a}^2\left(5-7x^2\right), \\
    \mathcal{A}_{7/2}^{S3} =& \tilde{a}^3\left(8x - 12x^3\right), \\
    \mathcal{A}_{4}^{S4} =& \frac{3}{16}\tilde{a}^4\left(1-x^2\right)\left(29-37x^2\right), \\
    \mathcal{A}_{11/2}^{S5} =& -\frac{3}{4}\tilde{a}^5x\left(1-x^2\right)\left(17-33x^2\right), \\
    \mathcal{A}_{6}^{S6} =& -\frac{1}{32}\tilde{a}^6\left(1-x^2\right)^2\left(307-417x^2\right), \\
    \mathcal{A}_{15/2}^{S7} =& \frac{1}{4}\tilde{a}^7x\left(1-x^2\right)^2\left(101-134x^2\right), \\
    \mathcal{A}_{8}^{S8} =& \frac{1}{2048}\tilde{a}^8\left(1-x^2\right)^2 \notag \\
    &\times\left(29587-76390x^2+37587x^4\right),
\end{align}
and in the angular momentum flux are
\begin{align}
    \mathcal{C}_{3/2}^{S1} =& \frac{3\tilde{a}}{x}-6\tilde{a}x,\\
    \mathcal{C}_{2}^{S2} =& \frac{1}{2}\tilde{a}^2\left(7-9x^2\right), \\
    \mathcal{C}_{7/2}^{S3} =& -\frac{1}{8}\tilde{a}^3\left(\frac{55}{x}-254x+223x^3\right), \\
    \mathcal{C}_{4}^{S4} =& \frac{1}{8}\tilde{a}^4(1-x^2)(47-83x^2), \\
    \mathcal{C}_{11/2}^{S5} =& \frac{1}{16}\tilde{a}^5(1-x^2)\left(\frac{155}{x}-1124x+1257x^3\right), \\
    \mathcal{C}_{6}^{S6} =& -\frac{1}{16}\tilde{a}^6\left(1-x^2\right)^2\left(123-313x^2\right), \\
    \mathcal{C}_{15/2}^{S7} =& -\frac{1}{1024}\tilde{a}^7(1-x^2)^2\bigg(\frac{12355}{x} \notag \\
    &-133062x+168835x^3\bigg), \\
    \mathcal{C}_{8}^{S8} =& \frac{1}{256}\tilde{a}^8(1-x^2)^2(2361-10882x^2 \notag \\
    &+7753x^4).
\end{align}

Just as we saw in the gravitational case \cite{CastETC25a}, the leading-spin terms in the scalar fluxes split into separate half-integer PN order sequences ($\coeF{(A/C)}{2n+3/2}^{S(2n+1)}$) and integer PN order sequences $(\coeF{(A/C)}{2n}^{S(2n)})$.  The former represent also the first appearances of odd powers of $a$ and the latter first appearances of even powers of $a$.  This structure is reminiscent of the leading-log sequences \cite{JohnMcDaShahWhit15,MunnEvan19a,MunnEvan20a} noted in gravitational fluxes from eccentric orbits.  It turned out in that problem that the leading-log enhancement functions in eccentricity had closed-form expressions that were predictable to all orders.  It remains to be seen whether the polynomials in $x$ in the leading-spin terms in this problem might also have a predictable pattern at all higher orders.  

The leading-spin terms in the scalar case are simpler than those in the gravitational case.  So a hunt for a pattern might plausibly start with those given here.  A cursory examination fails to reveal a predictable sequence.  However, the process in searching for the leading-log sequences \cite{MunnEvan19a,MunnEvan20a} was to start with how the first term in each sequence was constructed in PN theory alone from low-order source multiple moments.  The functional form of the first term yielded the required clues to predict higher-order terms.  To do the same for leading-spin sequences, we would need to shift attention back to the gravitational case and seek results from pure PN theory expansions for binaries with one spinning component, and make connection with the small mass ratio (perturbation theory) limit \cite{WitzETC25}.

\subsection{Comparison with Numerical Results}

For our expressions for the scalar fluxes, as with all PN expressions, they are valid for orbits with sufficiently large separations. However, being exact functions in spin $a$ and inclination $x$, they should be valid throughout the parameter space $\{a,x\}$. To validate and probe the applicability of our scalar flux expressions over the orbital parameter space, we compare our PN expanded scalar flux expressions with numerical scalar flux data generated from the \texttt{Teukolsky} package of the Black Hole Perturbation Toolkit \cite{TeukolskyBHPT,BHPTK18}. We evaluated the numerical Teukolsky solutions to sufficiently high precision (40 digits) such that numerical error in the solution is not visible on any of our plots.

\begin{figure*}[htb!]
    \includegraphics[width=\textwidth]{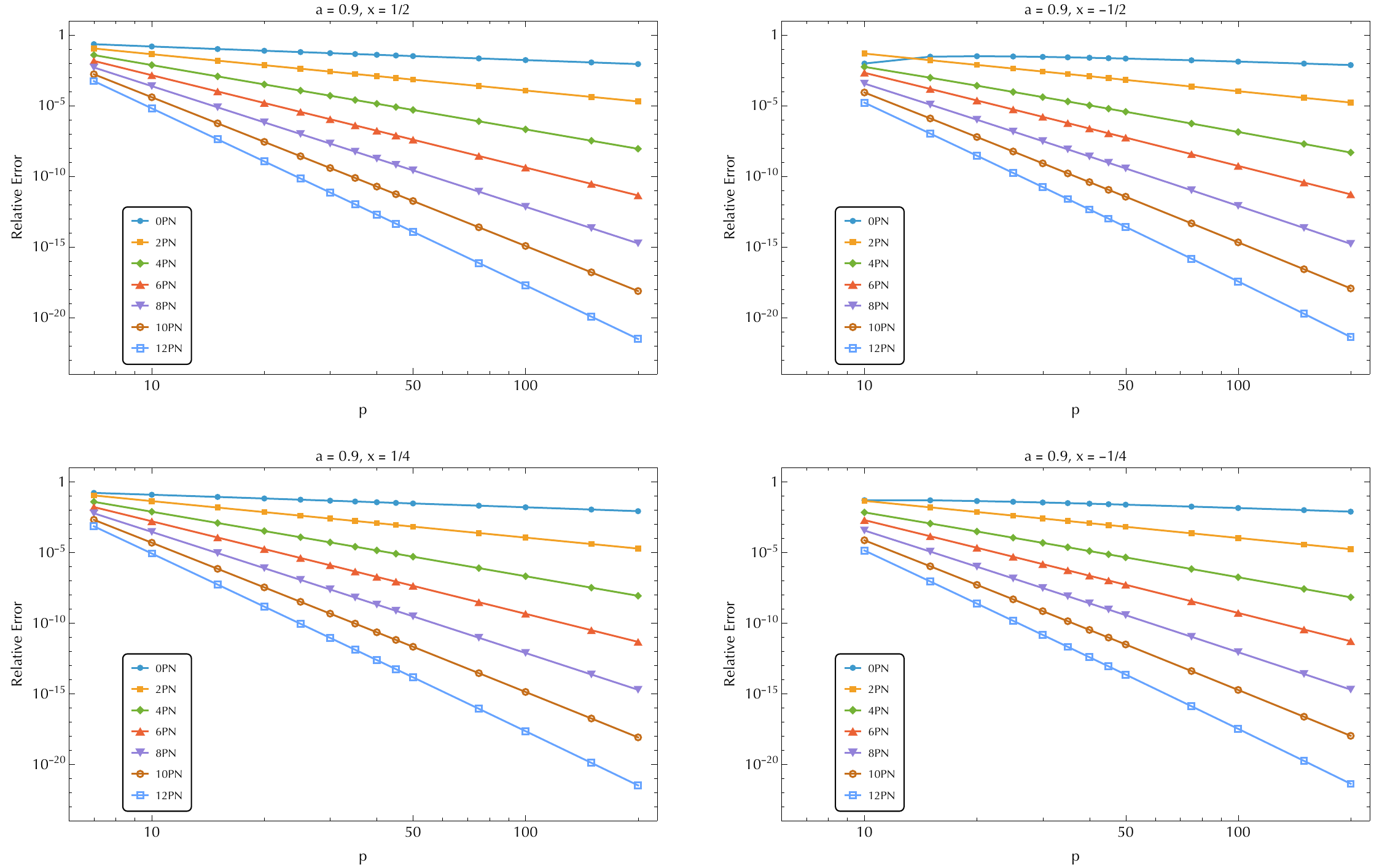}
    \caption{Relative error in the infinity energy flux when comparing the numerically evaluated scalar infinity flux expression with numerical scalar fluxes computed from a Teukolsky code. Results are shown as a function of $p$ for $a = 0.9M$ and $x \in \{1/2,-1/2,1/4,-1/4\}$. For retrograde orbits, $p = 7$ is excluded as the innermost stable spherical orbit for $a = 0.9M$ is located at $p_{ISSO} \simeq 7.1029$ \cite{SteiWarb19}. The residuals plotted in this figure are estimated to have a relative PN-scaling of $(n+1/2)$PN, where $n$ is the relative PN order of the flux expansion used.}
    \label{fig:NumComp}
\end{figure*}

In Figure \ref{fig:NumComp}, we present a comparison of the energy flux at infinity with $a = 0.9M$ and $x \in \{1/2,-1/2,1/4,-1/4\}$. We note that the fall-off becomes increasingly rapid as we include more terms in the expansion, with the expected residual scaling in $1/p$ for the number of PN terms included in the expansion. We also note that the fall-off behavior is similar for different values of $x$, although some of the low-PN fluxes from retrograde orbits differ significantly as the particle approaches the strong-field. We attribute this to our expansions being exact functions of $x$. 

\subsection{Near-extremal behavior}

In keeping $\atil$ arbitrary, we expect our scalar flux expressions to be valid through the entire parameter space. However, there are reasons to believe that the extremal limit, $\atil \to 0$, should be handled with care. For one, the degeneracy between the inner and outer horizons changes the nature of the singular points of the Teukolsky equation, with a modified MST method needed to compute the solutions \cite{CasaZimm19}. When combined with the comparison presented in Fig. \ref{fig:NumComp}, this leads to the following question: When does the expansion break down in the near-extremal regime? Here we perform an expansion in $\kappa \equiv \sqrt{1-\tilde{a}^2}$, and look at the resulting leading-order term. For the energy flux at infinity, we find that the first component with divergent-in-$\kappa$ behavior is
\begin{align}
    \coeF{A}{11} = -\frac{8192 \pi ^2 \left(1-x^2\right)^2}{27 \kappa }+O(\kappa^0)
\end{align}
This divergent-in-$\kappa$ behavior is significantly suppressed, appearing at 11PN at the earliest, then at 12PN. This suggests that the expansions in $\atil$ and $p^{-1}$ are noncommutative, and a separate expansion in the $\kappa \to 0$ limit should be considered. 

Before pressing forward with a separate near-extremal expansion, it is instructive to identify the sources of these divergent-in-$\kappa$ behaviors. It turns out that they originate from $m = 0$ modes, specifically terms involving the following gamma functions 
\begin{align}
    \Gam_{\tau}(z) = \Gam(1+f(\epsilon)+i\tau), 
\end{align}
where $f(\epsilon)$ is a polynomial function in $\epsilon$. In the $m \neq 0$ case, $\Gam_\tau$ is expanded about the point 
\begin{align}
    z_p = 1-\frac{ima}{\kappa},
\end{align}
which moves toward complex infinity as $\kappa \to 0$, and at the same time, the expansion point  
\begin{align}
    z-z_p = \frac{i\epsilon}{\kappa}+f(\epsilon),
\end{align}
widens at a slower rate when compared to the drift of $z_p$ towards complex infinity. Thus, for any value of $\kappa$, $\Gam_\tau$ for $m \neq 0$ is convergent. This changes when $m =0$, as $z_p = 1$, remaining fixed, thus for certain values of $\kappa$, $z-z_p$ will be outside the radius of convergence, $1>|z-z_p|$. This introduces the following constraint 
\begin{align}
    1 > \left|\frac{i\epsilon}{\kappa}+f(\epsilon)\right|,
\end{align}
suggesting an additional condition, $\epsilon < \kappa$, for convergence. In terms of $p$, we get the following relation
\begin{align}
    p > \kappa^{-2/3},
\end{align}
pushing the applicability of the scalar flux expressions towards orbits with larger separation. In practical terms, however, there are astrophysical limits on black hole spin, such as the Thorne limit, $a \simeq0.998M$ \cite{Thor74}, thus placing $p \gtrsim 6.3$ as a lower boundary of applicability for the flux expressions. This lower boundary is deep into the strong-field regime, where we expect the accuracy of our scalar flux expressions to falter.       

\section{Conclusion and Outlook}

In this work, we presented the expanded scalar flux to 12PN orders. The resulting PN-expanded expression is exact in spin $a$ and inclination parameter $x$. We highlight interesting features in the structure of the scalar fluxes, noting the expected asymmetric dependence of $x$ on $a$, as well as the presence of a complex relationship between $a$ and $x$ that begins to appear in the higher-order flux components. We also noted the behavior of the leading-spin terms in the scalar flux. Whether or not the leading-spin terms comprise a sequence function is the subject of a future study.

The PN-expanded scalar fluxes also showed excellent agreement with numerical scalar flux data. The expected exponential fall-off with increasing number of PN terms included is noted at each value of $p$, while getting noticeably worse as $p$ decreases. The performance of the analytical expressions presented here may be of interest for ongoing research on EMRIs as probes of the existence of scalar fields \cite{DellETC24}.

For future work, we are interested in extending our computation to the conservative sector, where regularized quantities local to the worldline are required. As a test run for the eventual gravitational case, we look to implement an analytical expansion of the regularized scalar field and related local quantities, with the specific goal of incorporating inclination to them. The scheme developed for the local scalar field will then be adapted for the gravitational case in a future work.  

\section*{Acknowledgements}

The authors thank Niels Warburton for providing numerical scalar flux data. The authors also thank Soichiro Isoyama for his helpful comments. This research was supported by NSF Grants PHY-2110335 and PHY-2409604 to the University of North Carolina at Chapel Hill and the Hamilton Award - University of North Carolina at Chapel Hill. CRE thanks UCD for support under UCD Seed Funding Grant 3057/SF2023. CK and JN acknowledge support from Science Foundation Ireland under Grant number 21/PATH-S/9610.

\appendix 

\begin{widetext}
    
\section{The Horizon Fluxes}

For the horizon fluxes, the PN structure can be written in a manner similar to the infinity flux, with the energy flux through the horizon written as
\begin{align}
    \begin{autobreak}
    \left\langle\frac{dE}{dt}\right\rangle_{\mathcal{H}} = 
    \frac{1}{3}q^2p^{-11/2}\bigg[ 
        \mathcal{B}_0
        +\mathcal{B}_1p^{-1}
        +\mathcal{B}_{3/2}p^{-3/2}
        +\mathcal{B}_{2}p^{-2}
        +\mathcal{B}_{5/2}p^{-5/2}
        +\left(\mathcal{B}_{3} + \mathcal{B}_{3L}\log(p)\right)p^{-3}
        +\mathcal{B}_{7/2}p^{-7/2}
        +\left(\mathcal{B}_{4}+\mathcal{B}_{4L}\log(p)\right)p^{-4}
        +\left(\mathcal{B}_{9/2}+\mathcal{B}_{9/2L}\log(p)\right)p^{-9/2}
        +\left(\mathcal{B}_{5}+\mathcal{B}_{5L}\log(p)\right)p^{-5}
        +\left(\mathcal{B}_{11/2}+\mathcal{B}_{11/2L}\log(p)\right)p^{-11/2}
        +\left(\mathcal{B}_{6}+\mathcal{B}_{6L}\log(p)+\mathcal{B}_{6L2}\log^2(p)\right)p^{-6}
    +\cdots\bigg],
    \end{autobreak}
\end{align} 
and the angular momentum flux through the horizon written as
\begin{align}
    \begin{autobreak}
        \left\langle\frac{dL_z}{dt}\right\rangle_{\mathcal{H}} = 
        \frac{1}{3}q^2xp^{-4}\bigg[
        \mathcal{D}_0
        +\mathcal{D}_1p^{-1}
        +\mathcal{D}_{3/2}p^{-3/2}
        +\mathcal{D}_{2}p^{-2}
        +\mathcal{D}_{5/2}p^{-5/2}
        +\left(\mathcal{D}_{3} + \mathcal{D}_{3L}\log(p)\right)p^{-3}
        +\mathcal{D}_{7/2}p^{-7/2}
        +\left(\mathcal{D}_{4}+\mathcal{D}_{4L}\log(p)\right)p^{-4}
        +\left(\mathcal{D}_{9/2}+\mathcal{D}_{9/2L}\log(p)\right)p^{-9/2}
        +\left(\mathcal{D}_{5}+\mathcal{D}_{5L}\log(p)\right)p^{-5}
        +\left(\mathcal{D}_{11/2}+\mathcal{D}_{11/2L}\log(p)\right)p^{-11/2}
        +\left(\mathcal{D}_{6}+\mathcal{D}_{6L}\log(p)+\mathcal{D}_{6L2}\log^2(p)\right)p^{-6}
        +\cdots\bigg].
    \end{autobreak}
\end{align}

We use the same decomposition used in Eqs. \eqref{eq:ECBD}-\eqref{eq:AMCBD} to separate the spin dependent components of $\mathcal{B}_{iLj}$ and $\mathcal{D}_{iLj}$, each written as 
\begin{align}
    \mathcal{B}_{iLj}(\tilde{a},x) = \mathcal{B}_{iLj}^{(0)} + \sum_{k=0}\mathcal{B}_{iLj}^{Sk}(\tilde{a},x), \qquad
    \mathcal{D}_{iLj}(\tilde{a},x) = \mathcal{D}_{iLj}^{(0)} + \sum_{k=0}\mathcal{D}_{iLj}^{Sk}(\tilde{a},x), 
\end{align}
with $\mathcal{B}_{iLj}^{(0)}$ and $\mathcal{D}_{iLj}^{(0)}$ representing the non-spinning limit of $\mathcal{B}_{iLj}$ and $\mathcal{D}_{iLj}$ respectively, while $\mathcal{B}_{iLj}^{Sk}$ and $\mathcal{D}_{iLj}^{Sk}$ are components proportional to $a^k$. 

The horizon-side flux expressions are far more complex and unwieldy to write down compared to the infinity-side flux expressions. Instead, we present the leading 2.5PN orders for illustration and refer to online repositories \cite{BHPTK18,UNCGrav22} for the full dataset, which is 10.5PN relative to the leading horizon-side flux or equivalently 12PN relative to the leading infinity-side flux. The scalar energy flux is 
\begin{align}
    \begin{autobreak}
        \MoveEqLeft
        \left\langle\frac{dE}{dt}\right\rangle_{\mathcal{H}} = 
        \frac{1}{3}q^2p^{-11/2}\bigg[-\atil x
        -2\atil xp^{-1}
        +\bigg(2(1+\kappa)
        -\atil^2\kappa(1-x^2)
        +5\atil^2x^2
        +a(1+x^2) \bar{\Psi}^{(0,1)}\left(\atil\right)\bigg)p^{-3/2}
        +\bigg(-3\atil x+2\atil^3x
        -3\atil^3x^3\bigg)p^{-2}
        +\bigg(4(1+\kappa)
        -3\atil^2
        -2\atil^2\kappa(1-x^2)
        +7\atil^2x^2
        +2\atil(1+x^2)\bar{\Psi} ^{(0,1)}(\atil)\bigg) p^{-5/2}
        +O\left(p^{-3}\right)\bigg],
    \end{autobreak}
\end{align}
while the scalar angular momentum flux is 
\begin{align}
    \begin{autobreak}
        \MoveEqLeft
        \left\langle\frac{dL_z}{dt}\right\rangle_{\mathcal{H}} =
        \frac{1}{3}q^2p^{-4}\bigg[-\left(\frac{\atil}{2}+\frac{\atil x^2}{2}\right)
        -\left(\atil+\atil x^2\right)p^{-1}
        +\left(2x+4 \atil^2x+2\kappa x+2\atil x \bar{\psi}^{(0,1)}(\atil)\right) p^{-3/2}
        -\left(\frac{3\atil}{2}-\frac{\atil^3}{16}+\frac{3 \atil x^2}{2}+\frac{\atil^3 x^2}{8}+\frac{15 \atil^3 x^4}{16}\right)p^{-2}
        +\bigg(4x
        +2\atil^2x
        +4\kappa x
        +4\atil x \bar{\Psi}^{(0,1)}(\atil)\bigg)p^{-5/2}
        +O\left(p^{-3}\right)\bigg].
    \end{autobreak}
\end{align}
We note that an analytical expression for scalar energy flux by a scalar particle in a circular orbit about a Reissner-N\"{o}rdstrom black hole has been previously computed up to 8.5PN \cite{BiniCarvGera16}. We find that the $a \to 0$ limit of our scalar energy flux expressions matches the Reissner-N\"{o}rdstrom flux expressions when the electric charge is zero.   

\begin{figure*}[htb!]
    \includegraphics[width=\textwidth]{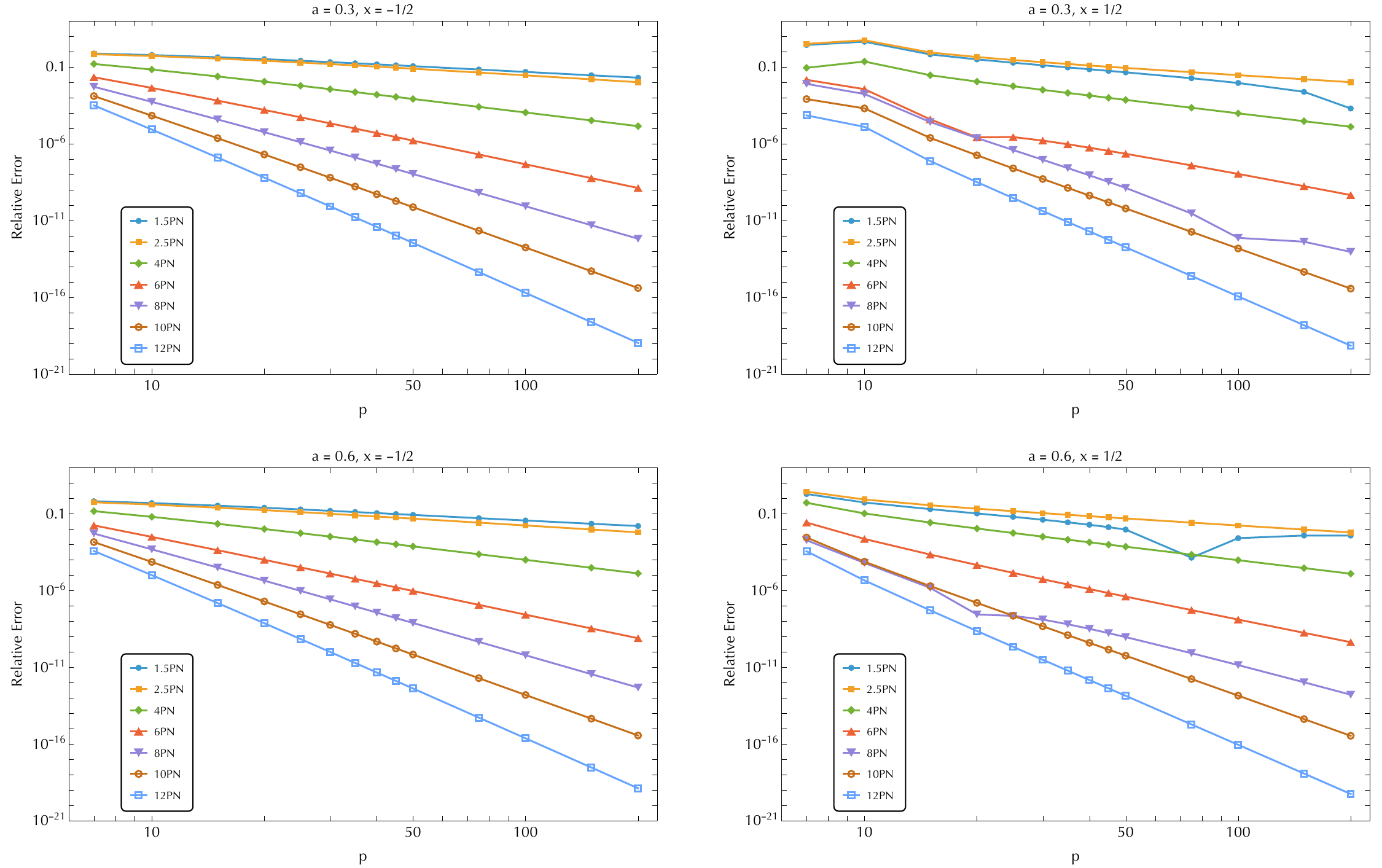}
    \caption{Relative error in the horizon energy flux when comparing the numerically evaluated scalar horizon flux expression with numerical scalar fluxes computed from a Teukolsky code. Results are shown as a function of $p$ for $a \in \{0.3M,0.6M\}$ and $x \in \{-1/2,1/2\}$. The residuals plotted in this figure are estimated to have a relative PN-scaling of $(n+1/2)$PN, where $n$ is the relative PN order of the flux expansion used. We note that several downward spikes appear in the log-of-absolute-value curves presented, these are due to a zero-crossing, $\lim_{p\to p_c}\Delta \dot{E}_{\mathcal{H}} = 0$, at that specific order of approximation.}
    \label{fig:horfluxErr}
\end{figure*}

In Fig. \ref{fig:horfluxErr} we present a comparison between the numerically evaluated PN expanded scalar horizon fluxes and the full numerical scalar horizon flux for orbits with $a \in \{0.3M,0.6M\}$ and $x \in \{-1/2,1/2\}$. Here we see the effects of changing $a$ and the orientation (prograde or retrograde) of the orbit, where same fall-off behavior observed from the infinity-side fluxes is present. We also note that zero-crossing behavior for the low PN contributions similar to the gravitational horizon flux \cite{CastETC25a,SagoFujiNaka25} are present.  

One final note, we also search for divergent in $\kappa$ behavior for the horizon fluxes. We find the lowest order component with divergent in $\kappa$ behavior is 
\begin{align}
    \coeF{B}{15/2}=\frac{16\pi^2(15-\pi^2)(1-x^2)}{135\kappa}+O(\kappa^0),
\end{align}
which is 7.5PN compared to the leading order horizon flux or 9PN compared to the leading order infinity flux. Stronger $\kappa^{-n}$ terms populate the flux components of higher PN order. As noted in Sec. IV.F, as a practical matter, these $\kappa^{-1}$ terms become relevant deep into the strong-field regime, where we do not expect applicability of our PN expansions.  

\end{widetext}
\bibliography{references}
\end{document}